\documentclass[english]{article}
\usepackage{amsmath}
\usepackage[T1]{fontenc}
\usepackage[utf8]{inputenc}
\usepackage{lmodern}
\usepackage[a4paper]{geometry}
\usepackage{babel}
\usepackage{authblk}
\usepackage{multicol}
\usepackage{geometry}
\usepackage[nottoc, notlof, notlot]{tocbibind}
\usepackage{graphicx}
\usepackage{float}
\usepackage{hyperref}
\usepackage{color,soul}
\usepackage{parskip}
\usepackage{}
\usepackage[squaren,Gray]{SIunits}

\title{Multi-scale friction coefficient : from roughness to system computation using deep learning}
\author[1]{Victor Lalleman}

\author[1]{Pierre Gosselet}
\author[2,3]{Cédric Hubert}
\author[4]{Stéphane Salengro}
\author[1]{Vincent Magnier}
\affil[1]{Univ. Lille, CNRS, Centrale Lille, UMR 9013 - LaMcube - Laboratoire de Mécanique, Multiphysique, Multiéchelle,
	F-59000 Lille, France, \url{victor.lalleman@univ-lille.fr}}
\affil[2]{UPHF, CNRS, UMR 8201 - LAMIH, Valenciennes, France}
\affil[3]{INSA Hauts-de-France, Valenciennes, France}
\affil[4]{MG-Valdunes, MA-Steel, France}
\date{\today}

\begin{document}

\maketitle

\begin{abstract}
	The presence of surface defects (roughness, surface imperfections, profiles, etc.) in a contact inevitably leads to the modification of its local properties, such as the coefficient of friction. In railway wheelsets, this surface condition is crucial as it dictates appropriate fatigue design for the final use. However, these local phenomena are not well understood and require a real step back. Therefore, the aim of this paper is to propose a multiscale numerical strategy to better understand these phenomena.

    The multiscale strategy is divided into two steps. Initially, an analysis by the Discrete Element Method (DEM) modelling the interaction of generated rough surfaces is carried out to determine the coefficient of friction. In a second step, the results of DEM are introduced into a structural calculation where the enrichment of the coefficient of friction is done on each finite element contact. Given the wide variety of potential surface defects (size, distribution, height, etc.), a large number of DEM simulations is performed. A specially developed deep learning program is then used to account for these dispersions. The application targeted in this paper is the fitting of a wheel on a railway axle.
	
\end{abstract}
\medskip


\newpage

\section{Introduction}

Axle-mounted rolling elements consist of two wheels, one axle shaft and other components such as bearings, bearing boxes, brake discs, or other necessary equipment for propelling motor vehicles. The environment and sustainable development require the integration of eco-design and better management of raw materials for the realization of these axles.

Research has shown that for axle-mounted systems, the entire manufacturing chain (forging + wheel fitting + cyclic loading) must be considered for a relevant design of the system in its typical usage \cite{Saad2016}. In order to predict the lifespan, numerical tools have been developed with a systemic perspective. However, based on experience, it has been demonstrated that there are local phenomena that influence the system's durability, highlighting the importance of considering a multiscale approach. More precisely, surface defects during the wheel fitting operation play a crucial role in the clamping between the wheel and the axle, and cyclic loading generates fretting-fatigue phenomena at the edges of the fitting surfaces \cite{yameo2004}. Modelling these phenomena is essential for the safety of the axles.

Determining the coefficient of friction is crucial in numerical analysis, especially when it comes to understanding and modelling the behaviour of materials in the presence of surface defects or heterogeneities. Surface defects, such as roughness or scratches, and material heterogeneities can significantly influence the friction between two contact surfaces, thus affecting the performance and reliability of mechanical components.

Numerical approaches, like the finite element method (FEM), are often used to simulate friction behaviour under these complex conditions. These simulations allow for the prediction of friction forces and the analysis of the impact of surface defects and heterogeneities on the overall system behaviour. Studies have shown that the accuracy of these numerical models strongly depends on the ability to correctly characterize surface properties and contact conditions.

Karupannasamy et al. introduced a multi-scale contact model for sheet metal forming processes, highlighting the significance of surface geometry of rough surfaces in predicting friction coefficients under varying nominal contact pressures \cite{KARUPANNASAMY2013222}.
The advent of computer modelling has also opened new perspectives in determining the coefficient of friction. Temizer and Wriggers propose a contact homogenization technique for extracting the macroscopic friction coefficient of a three-body friction system consisting of rigid particles embedded between an elastic solid with finite deformation and a rigid surface \cite{Temizer}. Moghaddam developed and validated a multiscale finite element model to analyse frictional interactions between the shoe and the floor, considering microscopic and macroscopic characteristics, and comparing model results with experimental data to assess its validity \cite{MOGHADDAM2018145}. Waddad considered roughness as normal stiffness at the contact and compared the evolution of contact pressure and temperature between a perfect model (meaning perfectly smooth surfaces) and a model with roughness \cite{Waddad}. Chaise developed semi-analytical methods to predict residual stresses and their effects on mechanical processes, considering the influence of friction, particularly in rolling, impacts, and ultrasonic peening, providing valuable tools for enhancing the lifespan of mechanical components \cite{Chaise}.Tribological tests, such as the ball-on-disc tribometer \cite{Passerat}, allow for a finer characterization of interactions between moving surfaces. This method applies to fields ranging from ball bearing manufacturing to mechanical component coatings.

 Numerical simulations based on methods like molecular dynamics enable the prediction of behaviour in complex systems at different scales \cite{Mollon}, from individual molecules to macroscopic structures. Kounoudji et al. analyse the issue of bolted joints by focusing on friction at the thread interfaces, using a discrete element method-based approach to better understand tribological interactions \cite{kounoudji}. Taboada and Renouf investigates the initiation and growth of a dry granular shear zone under seismic shearing and flash heating using a discrete element method \cite{Taboada}. Demonstrating its versatility, Iordanoff et al. illustrates how DEM can complement forming process studies by examining thermal effects in cutting processes, analysing subsurface damages during abrasion, and characterizing welding joints in Friction Stir Welding \cite{Iordanoff}. Hubert et al. uses Discrete Element Method and graph theory to simulate electrical conduction in continua, achieving satisfactory accuracy in predicting conduction and Joule heating in various domains, with potential applications in crack detection \cite{Hubert}. All these examples show the impact that microscopic effects can have on the coefficient of friction at higher scales. However, to the authors knowledge, there is never any direct generalization or transfer of information from the microscopic to the macroscopic scale.


DEM can be used to build fine-scale models, providing specific information, but suffers from relatively long computation times. With a view to generalization, one idea would be to combine a few DEM simulations, to be able to relate the surfaces in contact to the coefficient of friction on the scale of the simulation through the use of artificial intelligence.

Integrating artificial intelligence (AI) into scientific and technical fields has revolutionized the way we approach complex and multidimensional problems. With AI advancements, researchers and engineers can leverage machine learning algorithms capable of analyzing diverse datasets. Based on this data, AI models can identify subtle trends and nontrivial relationships. These models are used in tribology for segmentation and morphological analysis of wear/particle traces images \cite{Bouchot}, and to predict and understand tribological effects on system performance \cite{Motamedi}.

The interest in using AI to predict the coefficient of friction lies in its ability to significantly accelerate the characterization process. Instead of relying solely on often tedious repeated experiments, researchers can utilise AI models to simulate virtual interactions between different surfaces and conditions. Moreover, AI can help overcome limitations related to the availability of experimental data, intrapolating relationships between material properties and the coefficient of friction in less-studied situations.


The numerical chain creates a database from which AI is able to learn. The results from AI can then be directly introduced into the complete model of axle to enrich it. The result is a numerical model of the wheel-axle fitting operation that takes into account the roughness of the surfaces in contact. Hendrycks proposed numerical models for a better understanding of the fatigue behaviour of railway wheel-axle assemblies, including simulation of wheel seating on axle bearings, multiaxial cyclic loading on full-scale wheel/axle specimens, and fatigue life analysis using the cycle jump method, revealing dependencies on material behaviour and local refinement \cite{Hendrycks}. Yameogo investigates the fatigue behaviour of wheel-axle assemblies, focusing on fretting fatigue cracking beneath seating loads, employing finite element modelling to analyse residual stresses from press fitting and their redistribution under cyclic loading, validating classic fatigue criteria such as the Dang Van criterion for predicting fretting fatigue crack initiation and modelling crack propagation using the Paris law \cite{yameo2004}.

In industry, the wheel is inserted and moved using a press. A fitting curve plots the force required to move the wheel during the operation, and is used to check that the process runs smoothly. In this section, we will try to obtain a curve similar to that obtained with the initial model, and compare the stress fields within the two parts between the model that does not take roughness into account and the enriched model.

\begin{figure}[H]
	\centering
	\includegraphics[width=.9\columnwidth]{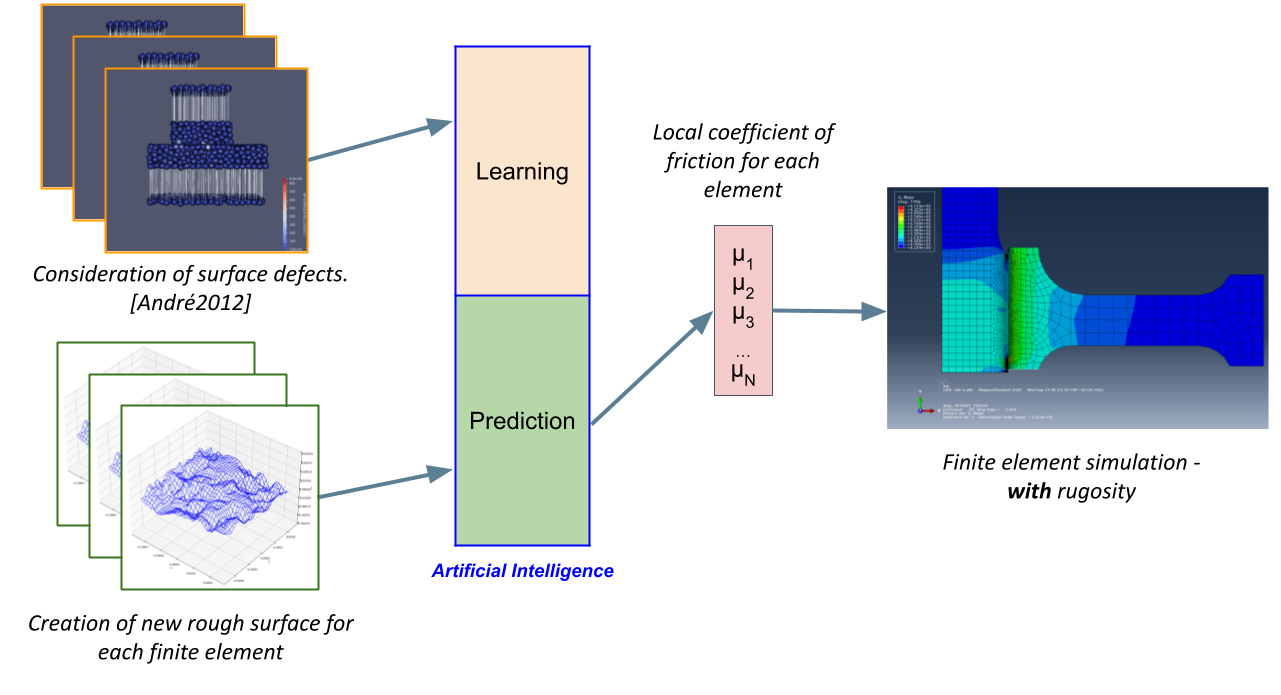}
	\caption{Multiscale modelling}
	\label{Multiscale calcul}
\end{figure}

In this paper, we conduct a new mesoscale modelling of the fitting operation using the discrete element method. The goal is to visualize the impact of surface defects (shape imperfections, roughness, etc.) on the evolution of the friction coefficient during the fitting operation, to generalize this to any type of surface by using an AI and to introduce these results into the finite element model of the fitting operation. The methodology is summarized in the Figure~\ref{Multiscale calcul}.

This paper is thus divided into 3 parts. In the first chapter, we are interested in the development of the DEM model simulating the fitting operation and in the interpretation of the results. In a second part, we develop a Deep-Learning artificial intelligence model capable of predicting from the database generated previously the mesoscopic coefficient of friction resulting from the sliding between two new surfaces. Finally in the last part, we introduce the results from the DEM model and generalized by AI in the FEM model and compare the results between the initial model in which roughness is not taken into account, with the enriched model.

\section{Multi-scale enrichment}

In parallel with the macroscopic finite element modelling of the fitting operation, we perform a second modelling of the operation at a more local scale using the discrete element method. This allows us to model surface defects experimentally observed on rough surfaces. The results are then fed into a deep learning artificial intelligence system to generalize the findings. The goal is to predict the friction coefficient resulting from the interaction of two new rough surfaces. Through this method, we generate a sufficient number of friction coefficients to incorporate them into the finite element model which is thus enriched by the local roughness.

\subsection{Macro-scale : Finite element model}

\begin{figure}[H]
	\centering
	\includegraphics[width=0.9\columnwidth]{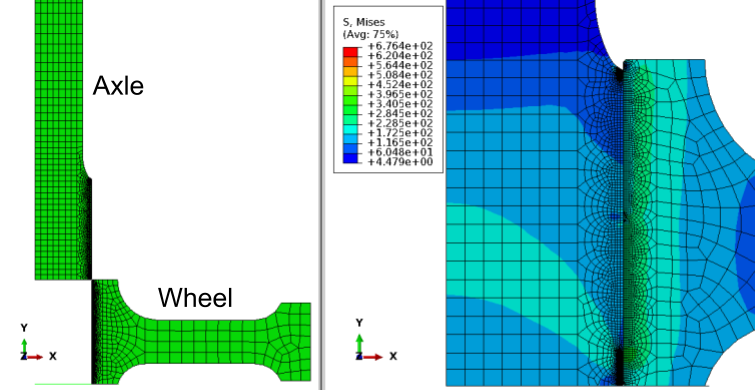}
	\caption{Finite element model of the wheel fitting operation on the axle}
	\label{FiniteElementModel}
\end{figure}

The macroscopic simulation of the wheel fitting operation is carried out using the finite element method as shown in Figure~\ref{FiniteElementModel}. The model is axisymmetric, and includes two domains with common axis: the axle and the wheel. A constant-speed displacement is applied to the underside of the wheel to slide it along the axle and position it. The clamping (difference between the outer radius of the axle and the inner radius of the wheel) is approximately 0.3 mm, and a contact with an overall coefficient of friction of 0.11 is imposed.

\subsection{Meso-scale : Discrete element model}

\subsubsection{Presentation of the model}

The use of the Discrete Element Method has enabled the modelling of the fitting operation conducted within the company at a more localized scale, namely the mesoscale, through the frictional modelling between two plates. 

\begin{figure}[H]
	\centering
	\includegraphics[width=0.6\columnwidth]{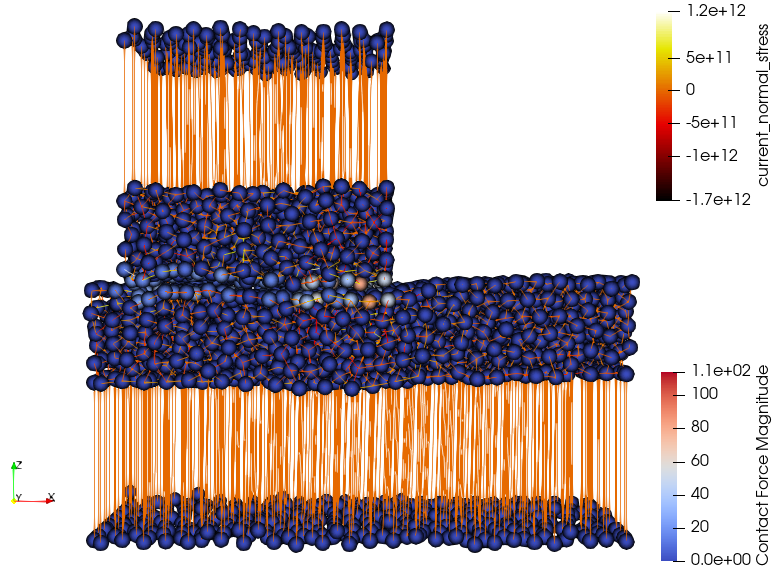}
	\caption{Discrete element model of the fitting operation at a local scale}
	\label{DiscreteElementModel}
\end{figure}

The upper plate which corresponds to the wheel is driven in motion while the  bottom plate which represents the axle remains stationary, see Figure~\ref{DiscreteElementModel}. With the aim of enhancing the existing finite element modelling, the domain size is set to match the size of the finite element, which is 1 mm.

In our case, we use a method called ``hybrid'' or ``lattice-beam''. This involves modelling matter as spheres, representing mass points, connected by Euler-Bernoulli beams. This allows us to model the elastic behaviour of materials and to obtain results on crack initiation and propagation within the model.

\subsubsection{Construction of the surface}

In order to create material domains composed of discrete elements (DE), we first define the external hull using a surface mesh. Then the DE software fills this envelope with particles and link them with beams.
Roughness is generated when the surface mesh is created, using the Diamond-Square algorithm \cite{Miller} which generates fractal surfaces. Only the faces in contact are affected by the roughness. The discrete elements adapt to this rough surface when the domain is filled, resulting in a compact domain. Characteristic roughness sizes are inspired by experimental measurements.

\begin{figure}[H]
	\centering
	\includegraphics[width=\columnwidth]{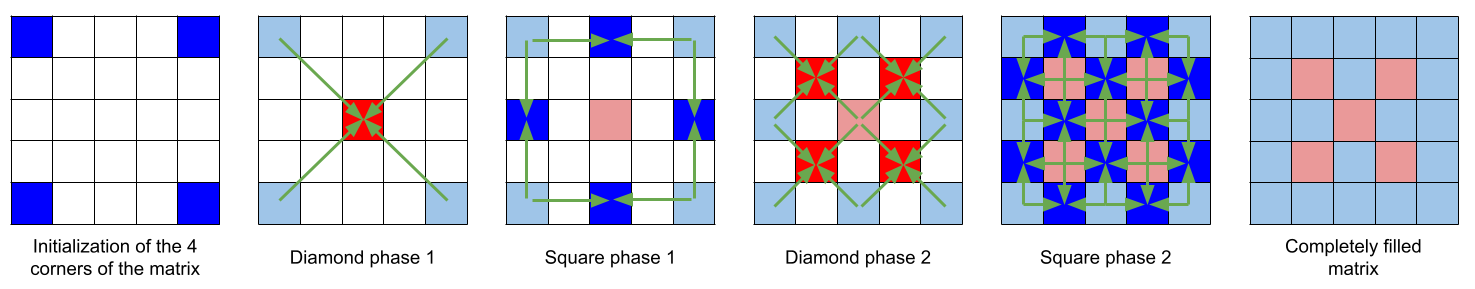}
	\caption{Steps of the Diamond-Square Algorithm}
	\label{DiamondSquare}
\end{figure}

The Diamond-Square algorithm involves defining a square matrix of size $2^{n}+1$, where $n$ is an integer. The height of the four corners of the matrix is initialized with random values within the interval [-$\frac{R}{2}$; $\frac{R}{2}$] where $R$ is the maximum height of defects observed on the axles and wheels. The matrix is then processed hierarchically, alternating between performing the diamond and square phases before reducing the step size by half, as described below and in Figure~\ref{DiamondSquare}:
\begin{enumerate}
	\item Diamond: At the center of each square, the average of the 4 points forming the square is calculated and then an additional random value is added to it.
	\item Square: At the center of each diamond, the average of the 4 points forming the diamond is calculated and then an additional random value is added to it.
	\item Finally, the step size is divided by two, and the process restarts from the Diamond step.
\end{enumerate}
Once the matrix is fully populated, the algorithm stops.

The location of the component in the matrix corresponds to the position $(x_{c},y_{c})$ on the surface, given that the points are uniformly distributed in both the $x$ and $y$ directions, and the value of the component corresponds to the height $z_c$ of the associated point. Thus, by using this algorithm twice, two clouds of points are obtained: one for the wheel and one for the axle.

\begin{figure}[H]
	\centering
	\includegraphics[width=.9\columnwidth]{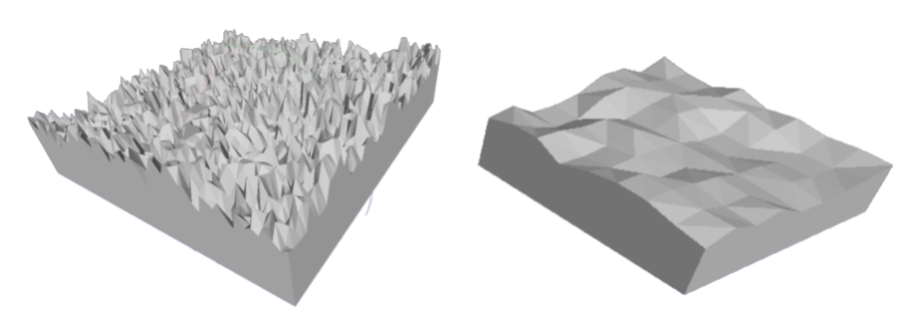}
	\caption{Examples of Surface Meshes. Left : n = 5 | Right : n = 3}
	\label{Surfaces}
\end{figure}

It is possible to generate a surface mesh that passes through all these points. To achieve this, triangular facets are defined with the vertices being the points obtained from the Diamond-Square algorithm. By proceeding with all the points and both point clouds, two surface meshes are obtained in the form of two boxes, with one of their faces being the rough surface. Furthermore, by varying the parameter $n$ in the Diamond-Square algorithm, it is possible to obtain more or fewer defects in our rough surfaces as shown in Figure~\ref{Surfaces}.

\begin{figure}[H]
	\centering
	\includegraphics[width=\columnwidth]{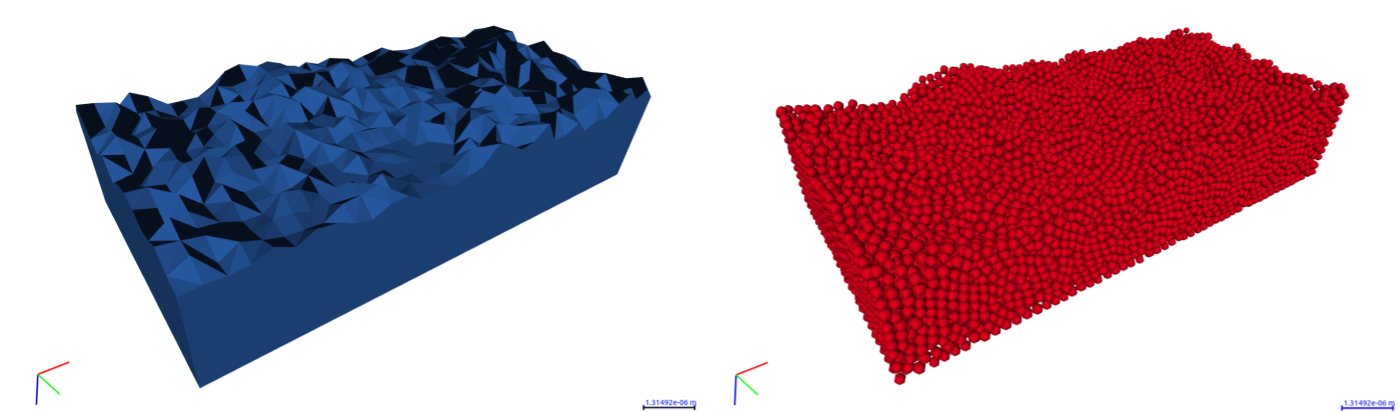}
	\caption{Example of a surface mesh obtained with the Diamond-Square algorithm, and the associated compact domain. Plate dimensions: 2mm x 1mm x 0.4mm}
	\label{CompactDomain}
\end{figure}

Finally, the two surface meshes are introduced into the GranOO-Cooker tool to fill them with paired discrete elements, see Figure~\ref{CompactDomain}. These clusters of particles form the two discrete domains. 

\subsubsection{Calibration of the microscopic parameters}\label{sub:calib}

To replicate the material behaviour in our model, it is necessary to quantify the interaction laws connecting the discrete elements. For this purpose, parameters need to be determined at the microscopic scale through tensile/compression test simulations. The method involves identifying microscopic parameters that yield the same macroscopic behaviour in the simulation as what is obtained experimentally.

In our case, the discrete elements are interconnected using cylindrical Euler-Bernoulli beams, allowing the geometry of the beam to be described with only two parameters: the length of the beam, which is automatically calculated by the software, and the radius of the beam. The beam radius is defined in GranOO through another parameter, the dimensionless radius:
\begin{equation}
    \tilde{r}_{\mu} = \frac{r_{\mu}}{(r_{1}+r_{2})/2}
\end{equation}
with:
\begin{itemize}
    \item $\tilde{r}_{\mu}$, the dimensionless radius of the cohesive beam;
    \item $r_{1}$, the radius of the first discrete element; 
    \item $r_{2}$, the radius of the second discrete element; 
    \item $r_{\mu}$, the radius of the cohesive beam; 
\end{itemize}

The elastic behaviour is defined by two parameters: the Young's modulus and Poisson's ratio. However, each microscopic parameter has varying degrees of influence on several macroscopic parameters. Therefore, it is necessary to establish the order in which the parameters are to be calibrated.

The method proposed by André et al. \cite{andre2012} consists in implementing a quasi-static numerical tensile test. Using the theory of strength of materials for Euler-Bernoulli beams, the macroscopic Young's modulus and Poisson's ratio can be determined analytically. By studying the impact of each microscopic parameter on the macroscopic ones, it is possible to develop a strategy for calibrating the numerical parameters.

The influence of the microscopic Poisson's ratio being negligible compared to the other parameters on the macroscopic behavior, we arbitrarily impose a Poisson's ratio value of 0,3. It then remains to study the influence of the microscopic Young's modulus and the radius ratio. For this, we study in a first case the influence of the radius of the beams on the macroscopic parameters, and those for several fixed values of microscopic modulus. Then we proceed in an analogous way to study the influence of the microscopic Young's modulus for several beam radius values.

\begin{figure}[H]
	\centering
	\includegraphics[width=\columnwidth]{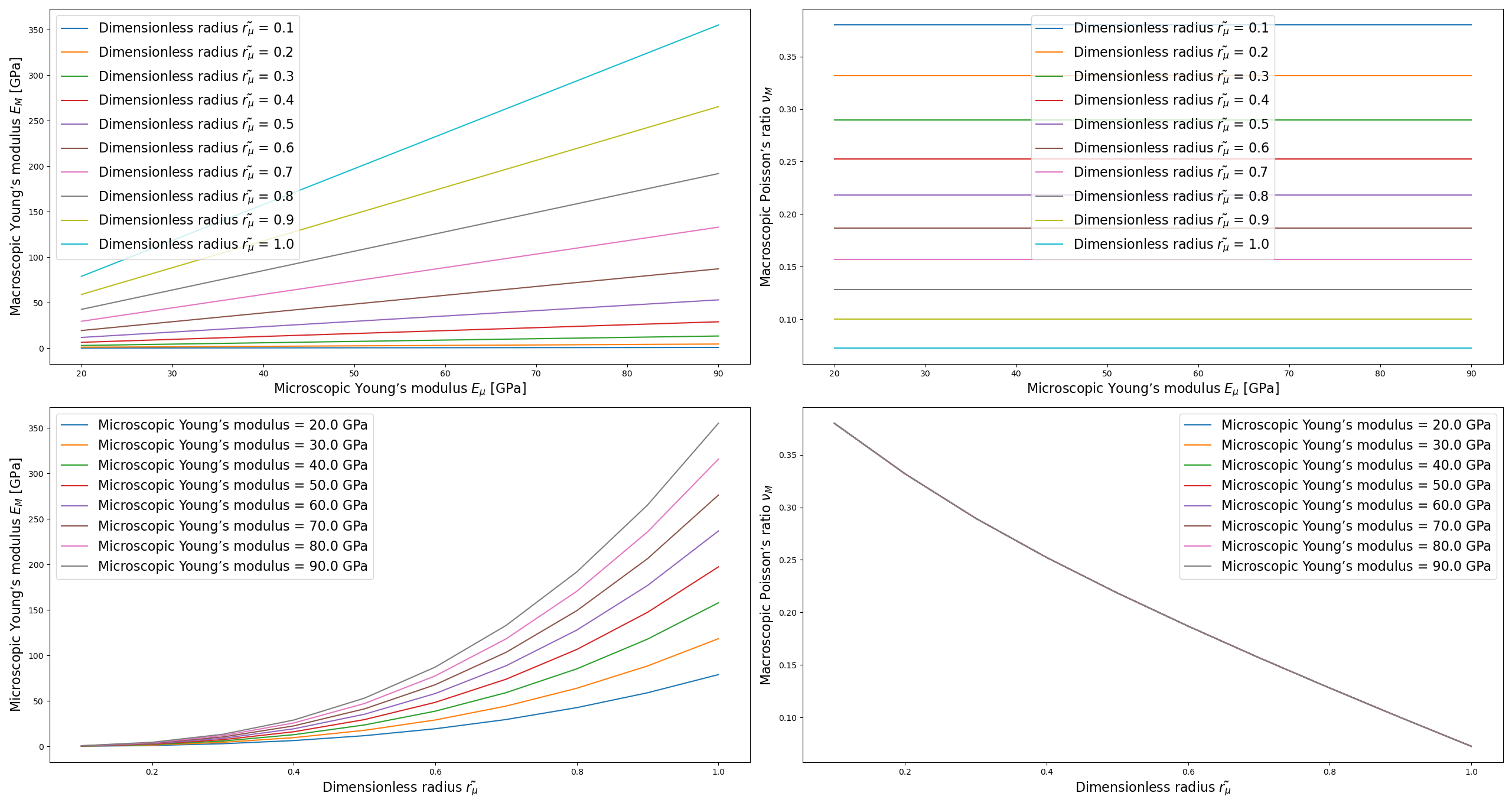}
	\caption{Influence of microscopic parameters on macroscopic parameters}
	\label{InfluenceParametresMicroscopiques}
\end{figure}

In view of the results on Figure~\ref{InfluenceParametresMicroscopiques}, the following strategy has been adopted:
\begin{enumerate}
    \item Arbitrary setting of the microscopic Poisson's ratio $\nu_{m}$ to 0.3. The variation of this parameter results in negligible change in macroscopic parameters (less than 1\%). The Poisson's ratio has a significant impact only in case of torsion. However, the beams experience very little torsion, primarily undergoing flexion and tension.
    \item Calibration of the microscopic dimensionless radius~$\tilde{r}_{\mu}$ with the macroscopic Poisson's ratio~$\nu_{M}$.
    \item Calibration of the microscopic Young's modulus~$E_{m}$ with the macroscopic Young's modulus~$E_{M}$.
\end{enumerate}

Finally, the density needs to be determined, given that mass is concentrated in the discrete elements. Due to the spherical shape of the discrete elements, void areas are present in our domains. The method involves increasing the density to compensate for these voids in order to ensure that the real and continuous domain has the same mass as the digital and discrete domain. The density is then determined by the following relationship:
\begin{equation}
    \rho_{\mu}=\frac{\rho_{M}V_{M}}{\sum\limits_{i=1}^{N}V_{\mu_{i}}}=\frac{\rho_{M}}{f}
\end{equation}
with:
\begin{itemize}
    \item $\rho_{\mu}$, the density of the discrete elements;
    \item $V_{\mu_{i}}$, the volume of the discrete element $i$;
    \item $\rho_{M}$, the macroscopic material density;
    \item $V_{M}$, the volume of the overall geometry;
    \item $f$, the volume fraction (approximately 0.64).
\end{itemize}

\subsubsection{Introduction of the stiffness of the domains}

\begin{figure}[H]
	\centering
	\includegraphics[width=0.5\columnwidth]{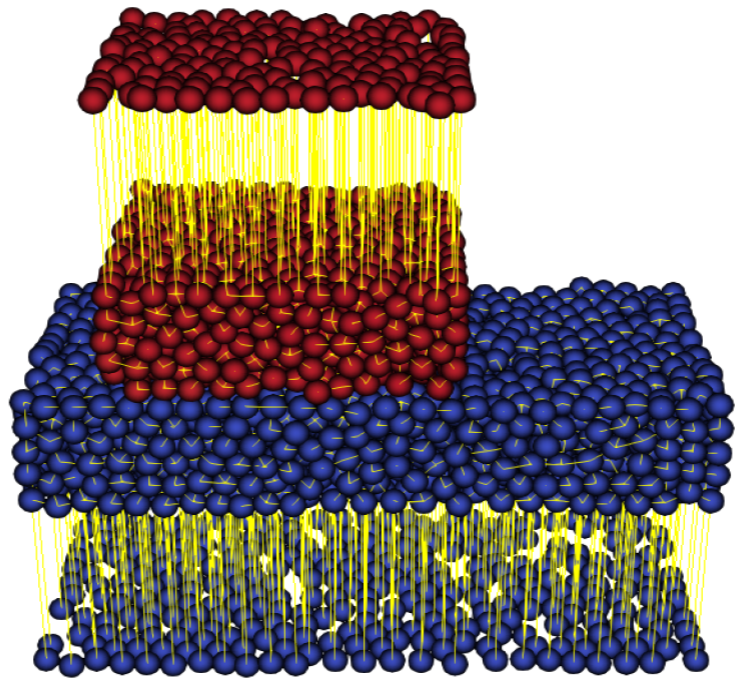}
	\caption{New Domains (Addition of Elements)}
	\label{NewDomains}
\end{figure}

Discrete Element computations are conducted in explicit dynamics with very small-time steps. They are extremely CPU consuming, and it is necessary to limit the number of particles so that computations do not take too long. In order to have sufficiently many particles in contact,  we have chosen slender geometries, (red and blue particle clusters on the figure~\ref{NewDomains}). It is crucial to define realistic boundary conditions on the top and bottom surfaces, that avoid the contact to be perturbed by edge effects propagating in the thickness. 

The boundary conditions (displacement control of the wheel and clamping of the axle) are applied to these new elements (which will be referred to as ``external faces''). This allows us to introduce flexibility into our model to achieve results closer to experimental observations.

Since we are adding new elements and beams, it's necessary to calibrate these new connections. While the beams within the two plates required modelling a tension test and using a bisection method to determine microscopic parameters (requiring a significant amount of computation time), here it is possible to directly calculate the radius $R$ and Young's modulus $E_m$ of the new beams analytically.

Regarding the radius, in order to prevent the rotation of the new beams it has been decided to use the radius of the discrete elements.

\begin{figure}[H]
	\centering
	\includegraphics[width=0.5\columnwidth]{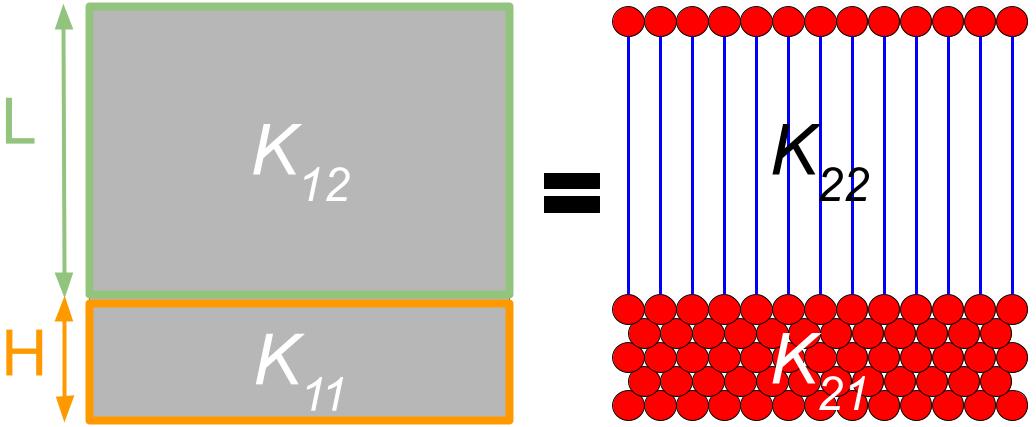}
	\caption{Analogy between the Real Domain and the Discrete Elements Domain}
	\label{Analogie}
\end{figure}

As for the Young's modulus, it's a matter of analogy between a cubic domain made of steel with a volume of $V = L^{3} = 1$ \milli\cubic\meter~and our domain composed of discrete elements, as shown in Figure~\ref{Analogie}. We consider that the cubic domain can be modeled as two equivalent springs $K_{11}$ and $K_{12}$ in series, and the discrete elements model can also be divided into two equivalent springs $K_{21}$ and $K_{22}$ connected in series. The springs with stiffness $K_{11}$ and $K_{21}$ have a length of $H$, which is the thickness of our plate made of discrete elements, while the other two springs have a length of $L-H$.

The DE model's parameters were calibrated in Subsection~\ref{sub:calib} in order to achieve a given macro stiffness, so that $K_{21}=K_{11}$. We need the stiffness $K_{22}$, constituted by a set of $N$ beams in parallel, to be equal to the stiffness $K_{12}$ of a cube with macroscopic Young's modulus $E_M$. This results in:
\begin{equation}
    E_{m} = \frac{E_{M} (L-H)^{2}}{N \pi R^{2}}
\end{equation}

\subsection{Results of the DE simulation and generalization to generate the database}

Once the DE model calibrated, simulations can be conducted. The displacement is imposed on the external faces. The axle is supposed to be clamped whereas the wheel's displacement is composed of two parts: the first part of the simulation involves bringing the wheel into contact with the axle and imposing a squeezing consistent with the sizing of the assembly, whereas the second part models the movement of the wheel along the axle. One last parameter needs to be provided: the microscopic coefficient of friction which models the interaction between particles in contact. Its value is calibrated so that the computed tangential force matches the experimental results.

\begin{figure}[H]
	\centering
	\includegraphics[width=1\columnwidth]{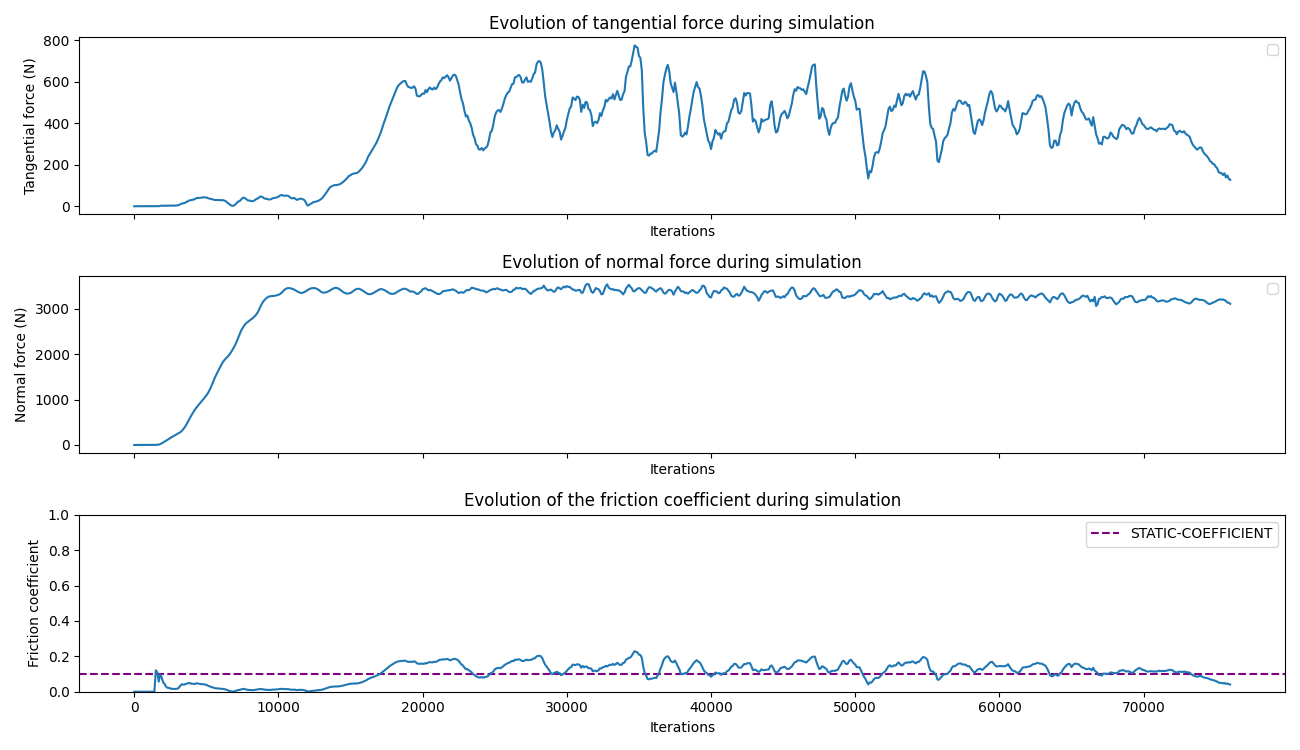}
	\caption{Evolution of normal and tangential forces, and the overall coefficient of friction}
	\label{Courbes}
\end{figure}

Among the possible outputs, the simulation provides the resulting forces in the external faces. By taking the ratio of these the tangential and normal parts, one can evaluate a coefficient of friction during the simulation as shown in Figure~\ref{Courbes}.

During the squeezing step, a nearly linear increase in both normal and tangential forces is observed, corresponding to the enlargement of the contact area between the wheel and the axle. In the second part, the overall normal force remains relatively constant due to constant clamping between the two plates. However, the tangential force exhibits several peaks, corresponding to interactions between elements on the axle and on the wheel. The smallest variations correspond to the meeting of two elements, while the largest are the result of the accumulation of several peaks and correspond to the meeting of two roughnesses composed of several discrete elements.

The entire computational process takes around 5 hours on average and provides information about the coefficient of friction over a $1 mm^{2}$ surface area. Considering the large number of simulations required to cover the entire contact surface in the finite element modelling, the use of artificial intelligence seems necessary. Specifically, a specialized branch of artificial intelligence known as Deep Learning will be employed, which falls under the umbrella of Machine Learning.

The files required for AI training are extracted from several discrete element simulations. The inputs include the surfaces in contact during the simulation, obtained by 3D interpolation of the position of the discrete elements, and the microscopic friction coefficient imposed in the simulation. The target is the coefficient of friction obtained by calculating the ratio of the tangential force to the normal force. These inputs and outputs are then grouped together in a database that is repeatedly processed by the AI to learn how to determine from the topology of two rough surfaces the coefficient of friction resulting from their sliding.

\section{Predicting the coefficient of friction using artificial intelligence}

\subsection{AI architecture}

As previously mentioned, the AI takes three inputs: the two surfaces in the form of two square matrices, corresponding to the wheel and axle, and the microscopic friction coefficient in the form of a real number. Input data are normalized between 0 and 1 to ensure stability, convergence and performance of the AI model by reducing scaling differences between features. The AI must then return, from this information, a mesoscopic friction coefficient value. Since the coefficient of friction is a real value, and evolves over a continuous interval, the AI's objective is to perform a regression. The AI model is developed here with the Keras library, in Python.

\begin{figure}[H]
	\centering
	\includegraphics[width=\columnwidth]{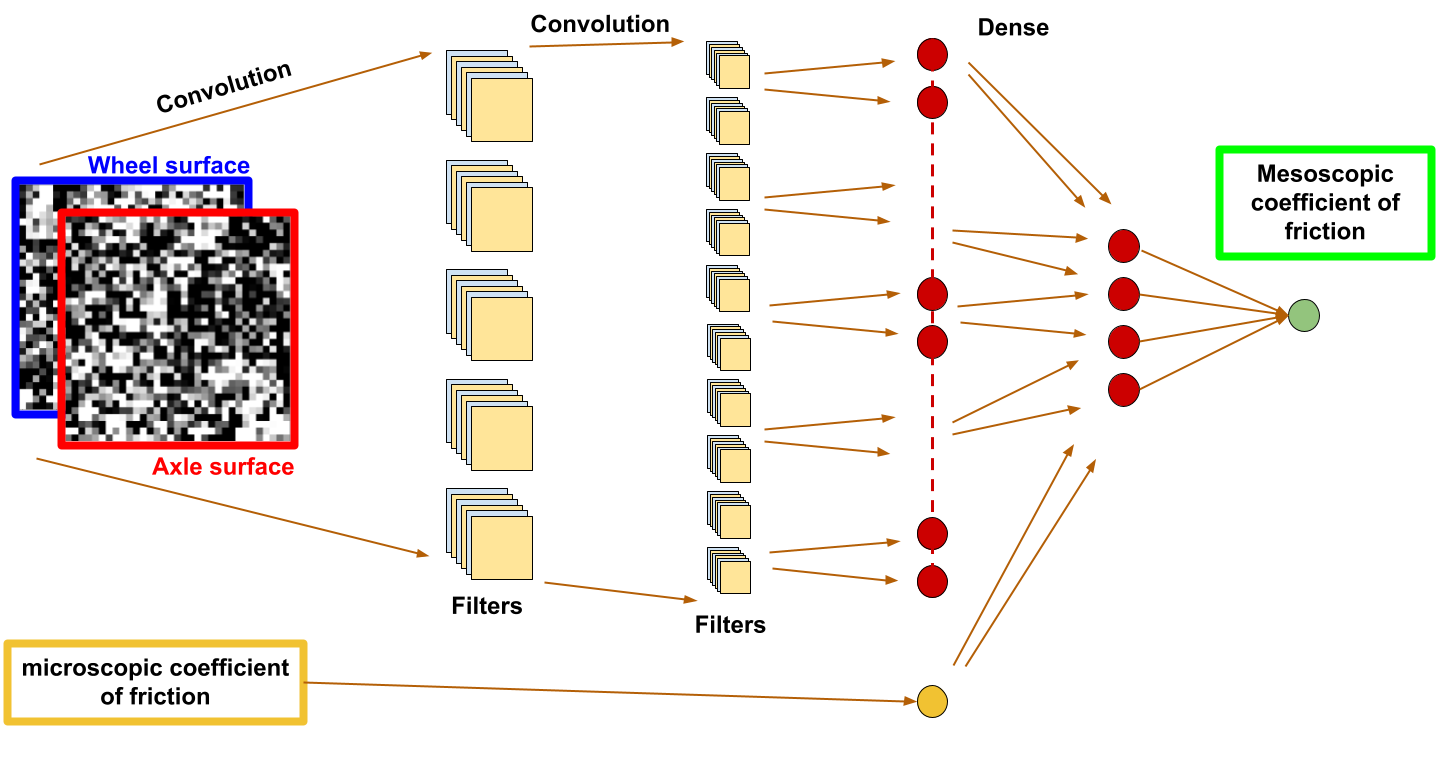}
	\caption{AI architecture}
	\label{ArchitectureIA}
\end{figure}

The model structure is composed of three branches and illustrated on Figure~\ref{ArchitectureIA}. The first two branches receive one of the surfaces to perform five 2D convolutions in order to consider spatiality, which is an essential element for results, with a \textit{relu} activation function which is consistent in the sense that there may or may not be contact. The number of filters is progressively decreasing, until data is vectorized with a \textit{flatten} layer. The third branch receives the microscopic friction coefficient.

The outputs of the three layers are merged into one vector with a \textit{concatenate} layer, which passes through five \textit{dense} layers with a \textit{relu} activation function. The optimizer used is \textit{adam} and the cost function used is \textit{mean\_squared\_error}.

\subsection{Model validation}

\begin{figure}[H]
	\centering
	\includegraphics[width=0.7\columnwidth]{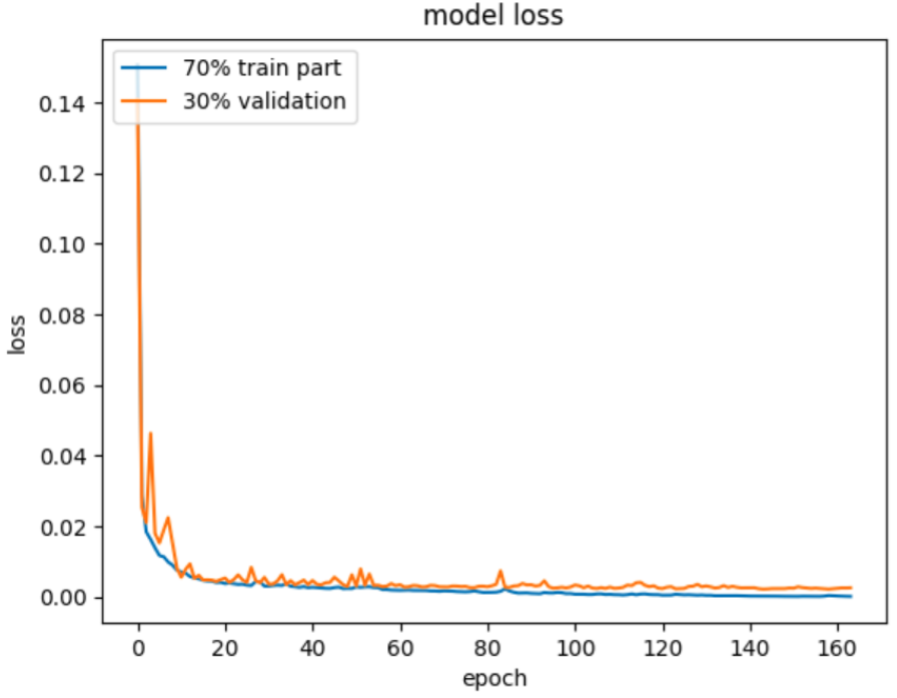}
	\caption{Evolution of Accuracy During Training}
	\label{CourbesLoss}
\end{figure}

In the definition of the model, a loss function is selected to assess the performance of the network during its operation. With each iteration of our dataset through the model (or epoch), it self-evaluates by measuring the difference (or loss) between the predicted outcomes and the actual results from our dataset. This allows us to gain insights into the behaviour of our model throughout the learning process, both on the training and validation datasets. The evolution of accuracy during training can then be plotted and visualized as two curves: a training curve and a validation curve. Learning takes is made with 70\% of the database, and validation with the remaining 30\%. In Figure~\ref{CourbesLoss}, both curves tend towards 0, indicating that the AI has been successfully trained.

\begin{figure}[H]
	\centering
	\includegraphics[width=0.9\columnwidth]{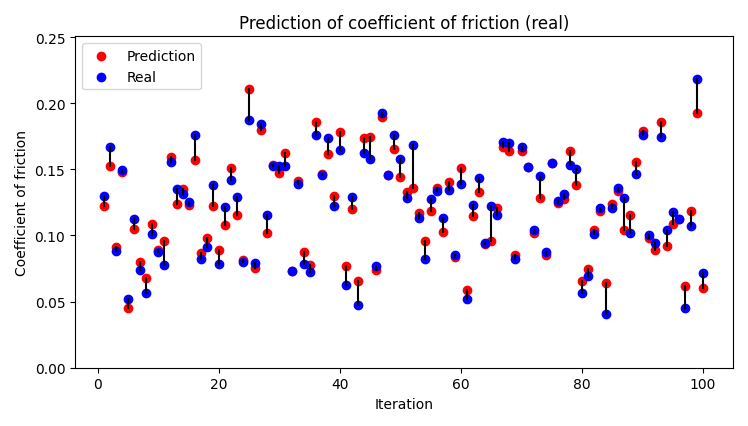}
	\caption{Comparison of actual results (blue) and predicted results (red)}
	\label{Prediction}
\end{figure}

As a second validation, the AI is tested on a new database that was not used during the training/validation and the coefficients of friction obtained with the DEM are compared with those predicted by the AI. Figure~\ref{Prediction} shows the values returned by the AI for 100 configurations. An error rate of 6.31\% is obtained, which is relatively low and which demonstrates the success of our model in generalizing the results. Moreover, predicting the coefficients of friction only took a few seconds, compared to a few hours for the discrete element simulation. With predictions being sufficiently accurate, the model can be used on new data and incorporate the results into the finite element simulation.

By comparing the values predicted by the AI with the actual values, we can clearly see how effective the model is at predicting the coefficient of friction. The biggest discrepancies are mainly observed for low coefficient values. This means that the database from which the AI learns does not contain enough configurations for which we obtain a low coefficient of friction. However, since these values represent only a small proportion of the total values obtained via DEM, and given the number of coefficients to be generated for a finite element simulation and the overall good accuracy of our model, we can neglect these errors.

\section{Integration of the results into the finite element model}

Two studies are carried out here to assess the impact of the microscopic coefficient of friction on the fitting curve, and that of the surfaces in contact on the stress field in the two parts.

\subsection{Influence of the microscopic coefficient of friction}

First, we generate enough surfaces to cover the entire calibration range, and ask the AI to predict the mesoscopic friction coefficients 5 times, each time with a different microscopic friction coefficient (${\mu}_{micro} \in [0.01,0.03,0.05,0.07,0.09]$). The predicted coefficients are then fed into the finite element model. This enables us to assess the influence of this parameter on the calibration model. 

\begin{figure}[H]
	\centering
	\includegraphics[width=0.9\columnwidth]{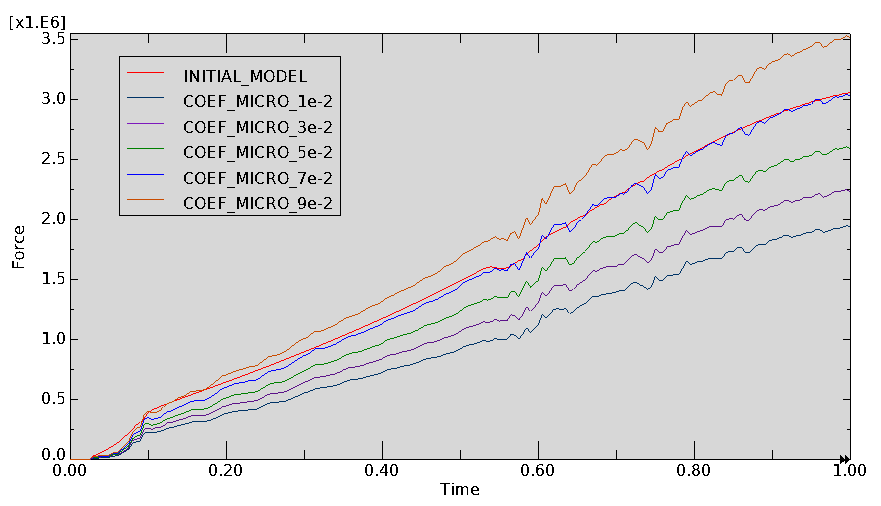}
	\caption{Evolution of the tangential force}
	\label{Etude_param}
\end{figure}

On Figure~\ref{Etude_param}, we plot the evolution of the tangential force at the contact surface of the axle in six different models: the initial model with a global coefficient of friction $\mu_{meso}=0.11$, and the five models obtained by predicting mesoscopic coefficients with different micro values.

If we begin by comparing the evolution of tangential force for the five enriched models, the surfaces being identical, we obtain the same curve profile. Only the amplitude varies, with the latter increasing with the mesoscopic friction coefficient.

\begin{figure}[H]
	\centering
	\includegraphics[width=\columnwidth]{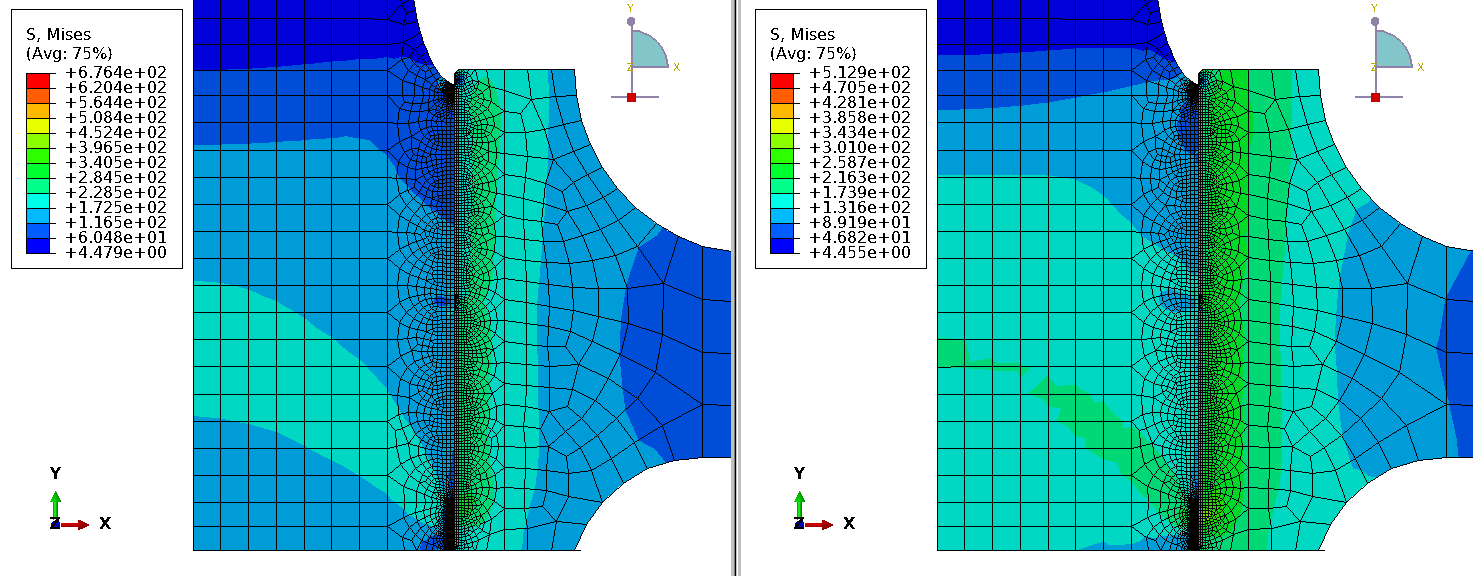}
	\caption{Stress field within the wheel and axle     a) Uniformed CoF     b) Enriched CoF}
	\label{StressField}
\end{figure}

\begin{figure}[H]
	\centering
	\includegraphics[width=\columnwidth]{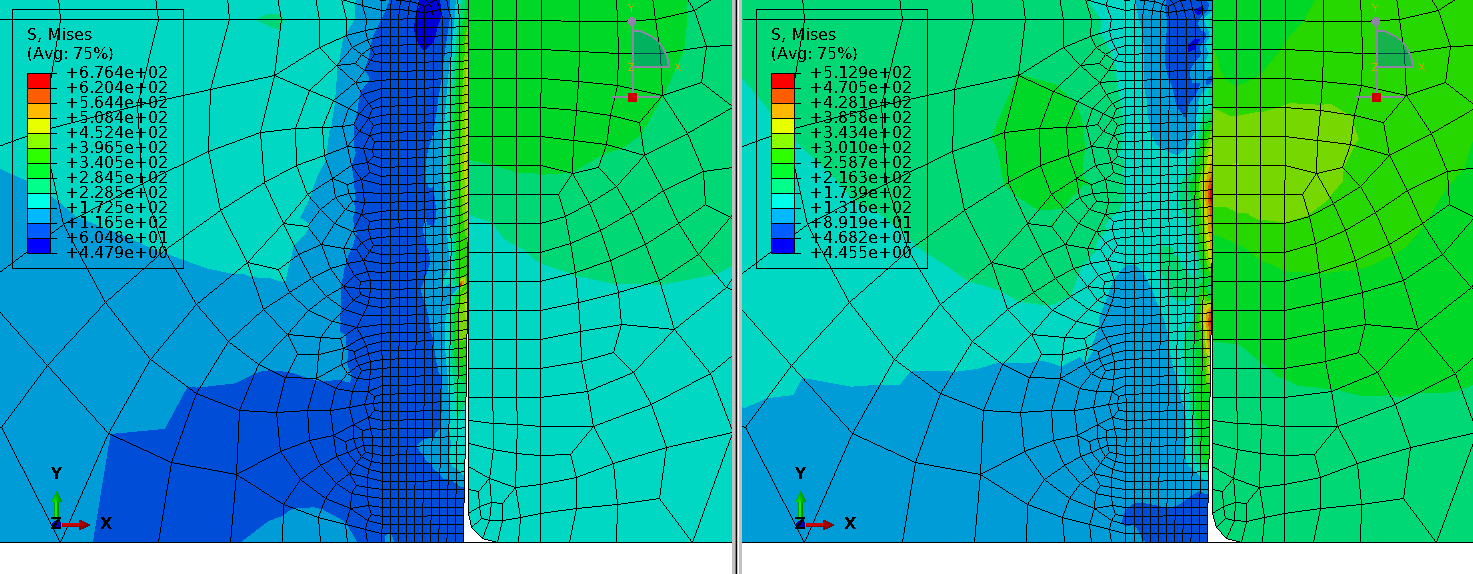}
	\caption{Zoom on the chamfer}
	\label{StressFieldZoom}
\end{figure}

Then, if we compare them with the evolution of effort obtained with the initial model (see Figures~\ref{StressField} and~\ref{StressFieldZoom}), we see that apart from a few variations due to the multiple changes of coefficients in the enriched model, one curve is quite close. This is the model obtained with the mesoscopic friction coefficient $\mu_{micro}=0.07$. We can then compare the stress fields obtained in the wheel/axle assembly with these two models.

In addition to the fact that the stress distribution differs between the two models, we note a difference in the maximum stress, localized in both cases at the entrance to the axle contact surface. In the first model, we obtain a maximum stress of 676.4 MPa, compared with 512.9 MPa in the enriched model, a difference of 25\%, which is far from negligible.

\subsection{Influence of the contacting surfaces}

In this section, we set the microscopic friction coefficient to 0.07, and compare the results between five models in which we generate new surfaces each time. 

\begin{figure}[H]
	\centering
	\includegraphics[width=0.9\columnwidth]{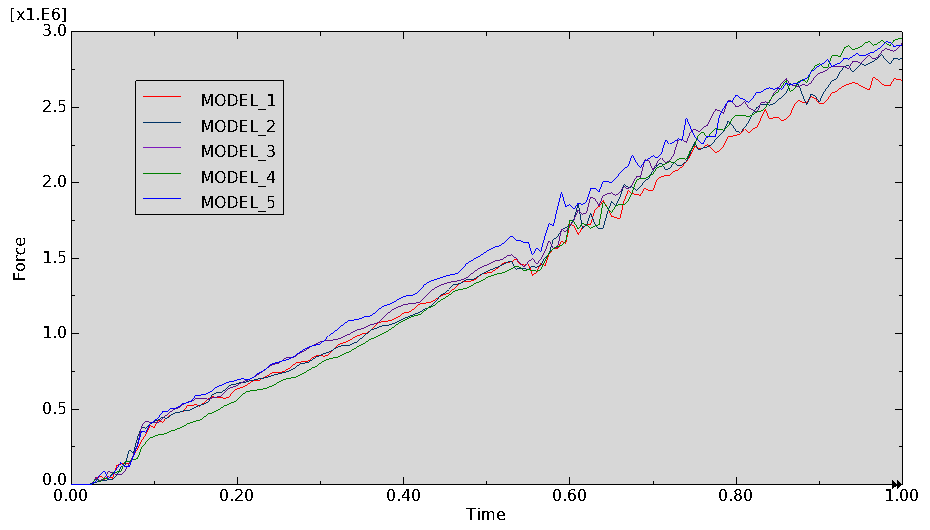}
	\caption{Evolution of the tangential force}
	\label{TangentielForceDifferentSurfaces}
\end{figure}

As before, we begin by comparing the evolution of tangential force between the 5 enriched models. The curves are printed on Figure~\ref{TangentielForceDifferentSurfaces}. We can see that for the same microscopic friction coefficient, we obtain quite different force evolutions for different surfaces. Moreover, the maximum value differs between the 5 models, ranging from $2.678 \times 10^{6}$ to $2.908 \times 10^{6}$ N.

\begin{figure}[H]
	\centering
	\includegraphics[width=\columnwidth]{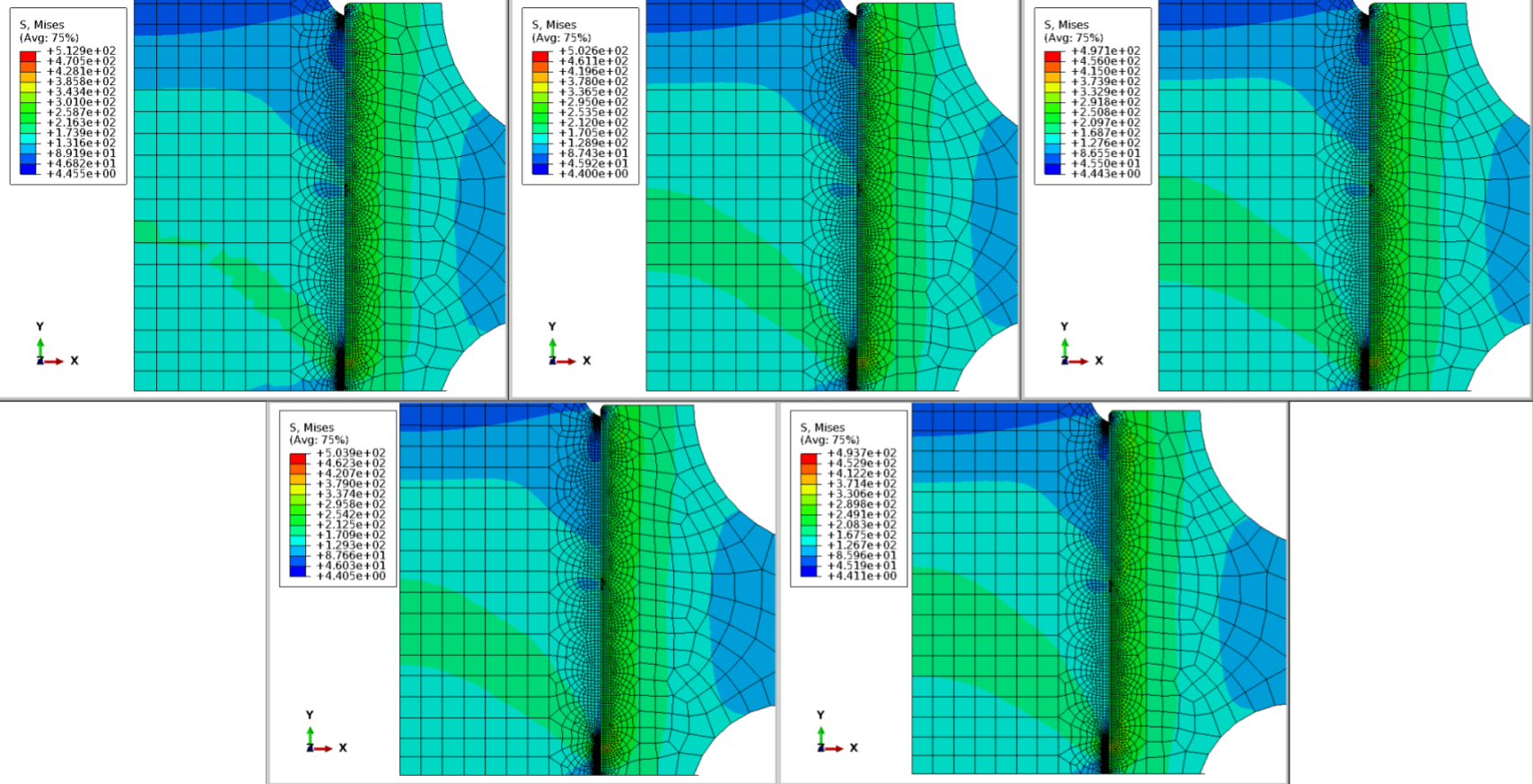}
	\caption{Stress field within the wheel and axle}
	\label{FieldDifferentSurfaces}
\end{figure}

Finally, we compare the constraint fields that are visible in Figure~\ref{FieldDifferentSurfaces}. Once again, we observe some differences due to higher or lower values of mesoscopic friction coefficients. The maximum stress, however, is always located at the chamfer entrance, but differs in all models and varies between 493.7 and 512.9 MPa. So, for different surfaces with the same geometrical characteristics, we obtain different stress values. This highlights the influence of rough surfaces on the overall behavior of the system, and the need to take them into account in a numerical model.

\section{Conclusions}
A multi-scale contact model has been developed, taking into account the surface state at fine scales. Two calculation methods (DEM and FEM) are interwoven with an enrichment of the friction coefficient. From a local point of view, a DEM calculation taking into account surface defects on a reduced size provides the evolution of normal and tangential forces. This information is fed back to the FEM calculation at element scale, via a representative friction coefficient. Generalization to various topologies has led us to use artificial intelligence, and more specifically deep learning. The latter takes into account interacting surfaces and the local coefficient of friction. All the models developed have been validated, confirming the robustness of the overall approach. An example is given of a wheel set-up on an axle, where it is shown that, from a local point of view, the stresses are greatly perturbed. These modifications will have an undeniable impact on fretting-fatigue phenomena, which remains a major cause of wheelset failure. Following on from this work, it remains to consider the source flow of surfaces, which will modify the coefficient of friction locally, and the evolution of the roughness, which will decrease with wear during the operation. This will require the creation of a new AI database with surfaces having different characteristics, and the contribution of experimental data to determine the evolution of rough surfaces during the operation.

\section*{Acknowledgement}
The present research work has been supported by the Hauts de France region, the MG-Valdunes company and the University of Lille.
Computations were conducted on the openstack cloud of the university of Lille.

\bibliographystyle{unsrt}
\bibliography{Bibliographie}

\begin{thebibliography}{10}

\bibitem{Saad2016}
Sofiane Saad, Vincent Magnier, Philippe Dufrenoy, Eric Charkaluk, and Fran{\c
  c}ois Demilly.
\newblock { Numerical Chain of Forging Railway Axle and Wheel Press Fitting
  Operation}.
\newblock In {\em {Sixth International Congress on Design and Modeling of
  Mechanical Systems, CMSM'2015}}, Hammamet, Tunisia, 2015.

\bibitem{yameo2004}
Arsène Yameogo.
\newblock {\em Etude expérimentale et numérique de l'amorçage et de la
  propagation de fissure de fretting dans un assemblage roue-essieu
  ferroviaire}.
\newblock PhD thesis, Ecole centrale de Paris 2004, 2004.
\newblock Thèse de doctorat dirigée par Prioul, Claude Mécanique et
  matériaux Châtenay-Malabry.

\bibitem{KARUPANNASAMY2013222}
D.K. Karupannasamy, M.B. {de Rooij}, and D.J. Schipper.
\newblock Multi-scale friction modelling for rough contacts under sliding
  conditions.
\newblock {\em Wear}, 308(1):222--231, 2013.

\bibitem{Temizer}
I~Temizer and Peter Wriggers.
\newblock A multiscale contact homogenization technique for the modeling of
  third bodies in the contact interface.
\newblock {\em Computer Methods in Applied Mechanics and Engineering},
  198(3-4):377--396, 2008.

\bibitem{MOGHADDAM2018145}
Seyed Reza~M. Moghaddam, Arjun Acharya, Mark~S. Redfern, and Kurt~E.
  Beschorner.
\newblock Predictive multiscale computational model of shoe-floor coefficient
  of friction.
\newblock {\em Journal of Biomechanics}, 66:145--152, 2018.

\bibitem{Waddad}
Yassine Waddad.
\newblock {\em Multiscale thermomechanical strategies for rough contact
  modeling: application to braking systems}.
\newblock PhD thesis, Lille 1, 2017.

\bibitem{Chaise}
Thibaut Chaise.
\newblock {\em {Mechanical simulation using a semi analytical method : from
  elasto-plastic rolling contact to multiple impacts}}.
\newblock Theses, {INSA de Lyon}, September 2011.

\bibitem{Passerat}
Stéphane Passerat.
\newblock {\em Comportement tribologique d'aciers et d'alliages de titane
  renforcés par traitements superficiels en glissement et en fretting}.
\newblock PhD thesis, Bordeaux 1 2000, 2000.
\newblock Thèse de doctorat dirigée par Denape, Jean Mécanique.

\bibitem{Mollon}
Komla~Ap{\'e}l{\'e}t{\'e} Kounoudji, Mathieu Renouf, Guilhem Mollon, and Yves
  Berthier.
\newblock Analyse tribologique d'un contact d'un assemblage boulonn{\'e} via la
  dem.
\newblock In {\em 12e Colloque national en calcul des structures}, 2015.

\bibitem{kounoudji}
Komla Kounoudji, Mathieu Renouf, Guilhem Mollon, and Yves Berthier.
\newblock Role of third body on bolted joints' self-loosening.
\newblock {\em Tribology Letters}, 61, 02 2016.

\bibitem{Taboada}
Alfredo Taboada and Mathieu Renouf.
\newblock {Rheology and breakdown energy of a shear zone undergoing flash
  heating in earthquake-like discrete element models}.
\newblock {\em {Geophysical Journal International}}, 233(2):1492--1514, 2023.

\bibitem{Iordanoff}
Ivan Iordanoff, Daniel Iliescu, Jean-Luc Charles, and J{\'e}rome N{\'e}auport.
\newblock {Discrete Element Method, a Tool to Investigate Complex Material
  Behaviour in Material Forming}.
\newblock In {\em {NUMIFORM 2010: Proceedings of the 10th International
  Conference on Numerical Methods in Industrial Forming Processes Dedicated to
  Professor O. C. Zienkiewicz (1921-2009)}}, pages p.778--786, South Korea,
  June 2010. {AIP Publishing}.

\bibitem{Hubert}
C{\'e}dric Hubert, Damien Andre, Laurent Dubar, Ivan Iordanoff, and Jean-Luc
  Charles.
\newblock {Simulation of continuum electrical conduction and Joule heating
  using DEM domains}.
\newblock {\em {International Journal for Numerical Methods in Engineering}},
  110(9):862--877, October 2016.

\bibitem{Bouchot}
Aliz{\'e}e Bouchot, Amandine Ferrieux-Paquet, Guilhem Mollon, Sylvie Descartes,
  and Johan Debayle.
\newblock Segmentation and morphological analysis of wear track/particles
  images using machine learning.
\newblock {\em Journal of Electronic Imaging}, 31(5):051605--051605, 2022.

\bibitem{Motamedi}
Nikzad Motamedi, Vincent Magnier, and Hazem Wannous.
\newblock Towards the identification of the link between the contact roughness
  and the friction-induced vibration: Use of deep learning.
\newblock {\em European Journal of Mechanics-A/Solids}, 99:104949, 2023.

\bibitem{Hendrycks}
Baptiste Hendrycks.
\newblock {\em Modélisation et optimisation multi-critères de la tenue en
  fatigue d'essieux ferroviaires innovants}.
\newblock PhD thesis, Lille 1, 2023.

\bibitem{Miller}
Gavin~SP Miller.
\newblock The definition and rendering of terrain maps.
\newblock In {\em Proceedings of the 13th annual conference on Computer
  graphics and interactive techniques}, pages 39--48, 1986.

\bibitem{andre2012}
Damien André, Ivan Iordanoff, Jean luc Charles, and Jérôme Néauport.
\newblock Discrete element method to simulate continuous material by using the
  cohesive beam model.
\newblock {\em Computer Methods in Applied Mechanics and Engineering},
  213-216:113--125, 2012.

\end{thebibliography}

\end{document}